\documentclass[10pt,superscriptaddress,prl,aps,twocolumn,nofootinbib,preprintnumbers]{revtex4}

\usepackage{float}
\usepackage{amsmath,amssymb}
\usepackage{soul}

\usepackage{graphicx}
\usepackage[caption=false]{subfig}

\usepackage{dcolumn}
\usepackage{bm}

\newcommand{\noplace}[1]{\affiliation{Void}}

\usepackage[colorlinks=true,linkcolor=blue,linktoc=all,citecolor=blue]{hyperref}
\usepackage[capitalise]{cleveref}
\usepackage{url}

\usepackage{breakurl}
\usepackage{amsmath}
\usepackage{graphicx} 
\usepackage{bm}
\usepackage{xcolor}

\begin{document}

\title{An extraction of the Collins–Soper kernel \\ from a joint analysis of experimental and lattice data}

\author{Artur Avkhadiev}
\thanks{Email: aavkhadi@anl.gov -- \href{https://orcid.org/0000-0003-3493-8649}{ORCID: 0000-0003-3493-8649}}
\affiliation{Center for Theoretical Physics\thinspace--\thinspace A Leinweber Institute, Massachusetts Institute of Technology, Cambridge, MA, USA 02139}
\affiliation{Physics Division, Argonne National Laboratory, Lemont, IL 60439, USA}%

\author{Valerio Bertone}
\thanks{E-mail: valerio.bertone@cea.fr -- \href{https://orcid.org/0000-0003-0148-0272}{ORCID: 0000-0003-0148-0272}}
\affiliation{IRFU, CEA, Universit\'e Paris-Saclay, F-91191 Gif-sur-Yvette, France}

\author{Chiara Bissolotti}
\thanks{E-mail: cbissolotti@anl.gov -- \href{https://orcid.org/0000-0003-3061-0144}{ORCID: 0000-0003-3061-0144}}
\affiliation{Argonne National Laboratory, PHY Division, Lemont, IL, USA}

\author{Matteo Cerutti}
\thanks{E-mail: matteo.cerutti@cea.fr -- \href{https://orcid.org/0000-0001-7238-5657}{ORCID: 0000-0001-7238-5657}}
\affiliation{IRFU, CEA, Universit\'e Paris-Saclay, F-91191 Gif-sur-Yvette, France}

\author{Yang~Fu}
\thanks{E-mail: yangfu@mit.edu --
\href{https://orcid.org/0000-0002-8965-1402}{ORCID: 0000-0002-8965-1402}}
\affiliation{Center for Theoretical Physics\thinspace--\thinspace A Leinweber Institute, Massachusetts Institute of Technology, Cambridge, MA, USA 02139}%

\author{Simone Rodini}
\thanks{E-mail: simone.rodini@unipv.it -- \href{https://orcid.org/0000-0002-8057-5597}{ORCID: 0000-0002-8057-5597}}
\affiliation{Deutsches Elektronen-Synchrotron DESY, Notkestr. 85, 22607 Hamburg, Germany}
\affiliation{Dipartimento di Fisica ``Alessandro Volta'', Universit\`a degli Studi di Pavia, 27100 Pavia, Italy}

\author{Phiala Shanahan}
\thanks{E-mail: phiala@mit.edu --
\href{https://orcid.org/0000-0002-0916-7603}{ORCID: 0000-0002-0916-7603}}
 \affiliation{Center for Theoretical Physics\thinspace--\thinspace A Leinweber Institute, Massachusetts Institute of Technology, Cambridge, MA, USA 02139}%
 
\author{Michael Wagman}
\thanks{E-mail: mwagman@fnal.gov -- \href{https://orcid.org/0000-0001-7670-1880}{ORCID: 0000-0001-7670-1880}}
  \affiliation{Fermi National Accelerator Laboratory, Batavia, IL 60510, USA}  
  
\author{Yong Zhao}%
\thanks{Email: yong.zhao@anl.gov -- \href{https://orcid.org/0000-0002-2688-6415}{ORCID: 0000-0002-2688-6415}}
 \affiliation{Physics Division, Argonne National Laboratory, Lemont, IL 60439, USA}

\begin{abstract}
We present a first joint extraction of the Collins–Soper kernel (CSK) combining experimental and lattice QCD data in the context of an analysis of transverse-momentum-dependent distributions (TMDs). Based on a neural-network parametrization, we perform a Bayesian reweighting of an existing fits of TMDs using lattice data, as well as a joint TMD fit to lattice and experimental data. We consistently find that the inclusion of lattice information shifts the central value of the CSK by approximately 10\% and reduces its uncertainty by 40-50\%, highlighting the potential of lattice inputs to improve TMD extractions.
\end{abstract}

\preprint{DESY-25-145; FERMILAB-PUB-25-0784-T; INT-PUB-25-026; MIT-CTP/5946}

\maketitle

\section{Introduction}

This work presents the first effort to incorporate lattice QCD data into a phenomenological extraction of transverse-momentum-dependent distributions (TMDs), to better constrain the Collins-Soper kernel (CSK). This shares the same idea as past combined analyses of lattice and experimental data towards a better understanding of the hadronic structure through the extraction of collinear distributions. These include analyses of unpolarized, longitudinally-polarized, and transversity parton distribution functions (PDFs)~\cite{Lin:2017stx,Cichy:2019ebf,DelDebbio:2020cbz,DelDebbio:2020rgv,Bringewatt:2020ixn,JeffersonLabAngularMomentumJAM:2022aix,Hou:2022onq,Karpie:2023nyg,Ablat:2024muy,Cocuzza:2025qvf,Ablat:2025xzm,Good:2025nny,Barry:2025wjx}, and generalized parton distributions~\cite{Guo:2023ahv,Cichy:2024afd,Guo:2025muf}. However, this is the first time that a similar strategy is employed in the context of a TMD extraction (see also Ref.~\cite{Cridge:2025wwo} for a similar effort in this direction). In this context, the CSK plays a special role in that it encodes information on the QCD vacuum rather than on the actual structure of hadrons. In fact, there is no experimental observable directly sensitive to CSK, so lattice is the only source of direct information. Improved determinations of the CSK are important to further sharpen control and reduce model-dependence in TMD evolution, enhance the precision of QCD predictions of multiple processes, and test the validity of factorization in the nonperturbative regime.

Formally, the CSK $K$ is the expectation value of a Wilson loop running along the light-front directions and characterized by a transverse displacement $\mathbf{b}_{\rm T}$. When $b\equiv|\mathbf{b}_{\rm T}|$ is small enough, the CSK can be computed in perturbation theory. As customary, radiative corrections introduce ultraviolet divergences, which, upon renormalization, lead to the introduction of the renormalization scale $\mu$. The dependence of $K$ on $\mu$ is governed by the renormalization-group equation:
\begin{equation}
\frac{\partial K(b,\mu)}{\partial\ln\mu}=-\gamma_{K}(\alpha_s(\mu))\,,
\label{eq:RGECSK}
\end{equation}
where $\gamma_K$, a.k.a. cusp anomalous dimension, is computable in perturbation theory and is currently fully known to four loops~\cite{Moult:2022xzt,Duhr:2022yyp}. Introducing the scale $\mu_b=2e^{-\gamma_{\rm E}}/b$, where $\gamma_{\rm E}$ is the Euler-Mascheroni constant, the solution to Eq.~(\ref{eq:RGECSK}) reads:
\begin{equation}
K(b,\mu)=K(b,\mu_b)-\int_{\mu_b}^\mu\frac{d\mu'}{\mu'}\gamma_K(\alpha_s(\mu'))\,,
\label{eq:RGECSKSol}
\end{equation}
where $K(b,\mu_b)$ is also known to four loops~\cite{Moult:2022xzt,Duhr:2022yyp}. When $b$ grows, such that $\mu_b\lesssim\Lambda_{\rm QCD}$, nonperturbative effects dominate and Eq.~(\ref{eq:RGECSKSol}) becomes unreliable. Nonperturbative affects are accounted for by modifying Eq.~(\ref{eq:RGECSKSol}) as follows:
\begin{equation}
K(b,\mu)=K(b_*,\mu_{b_*})-\int_{\mu_{b_*}}^\mu\frac{d\mu'}{\mu'}\gamma_K(\alpha_s(\mu'))-g_K(b)\,,
\label{eq:RGECSKSolNP}
\end{equation}
with $\mu_{b_*}=2e^{-\gamma_{\rm E}}/b_*$. Here, $b_*\equiv b_*(b)$ behaves linearly for small values of $b$ ($b_*\sim b$) and saturates to some $b_{\rm max}\ll\Lambda_{\rm QCD}^{-1}$ for large $b$. The purpose of $b_*$ is to guarantee that the first two terms in the r.h.s. of Eq.~(\ref{eq:RGECSKSolNP}) are evaluated in the perturbative regime where $\alpha_s\ll 1$. The role of the functions $g_K$ is to encode nonperturbative effects that arise from the large-$b$ region. A typical parametrization is:
\begin{equation}
g_K(b)=2g_2^2b^2\,,
\label{eq:gKparam}
\end{equation}
with the parameter $g_2$ to be determined from observables sensitive to the CSK. Other models for $g_K$, with at most two free parameters, have been used in recent TMD fits (see e.g. Refs.~\cite{Moos:2025sal,Aslan:2024nqg,Bacchetta:2019sam}). However, in this context, we found no evidence for the need of a more complicated functional form. 
The purpose of this work is therefore to determine the parameter $g_2$ from a joint analysis of experimental and lattice data.

\section{Baseline TMD fit}\label{sec:baseline}

A determination of $g_2$ from experimental data implies an extraction of TMDs. TMD analyses have achieved a remarkable level of sophistication: they are often based on broad, multiprocess data sets, they achieve high perturbative accuracy, and employ advanced methodological frameworks (see, e.g., Refs.~\cite{Bacchetta:2024qre, Bacchetta:2022awv, Moos:2023yfa, Moos:2025sal, Camarda:2022qdg, Barry:2025glq,Cuerpo:2025zde,Kang:2024dja,deFlorian:2004mp,Camarda:2025lbt}).
In this work, we rely on the TMD extraction presented in Ref.~\cite{Bacchetta:2025ara}. In that analysis, TMD parton distribution functions (PDFs) were extracted from a comprehensive set of Drell-Yan data using next-to-next-to-next-to-leading logarithmic (N$^3$LL) accurate predictions. The nonperturbative part of TMD PDFs, which also involves $g_2$, was parametrized through a neural network (NN). The flexibility of the NN enabled a solid estimate of the nonperturbative parameters and their uncertainties. That analysis obtained:
\begin{equation}
g_2^{\rm Baseline} = 0.186 \pm 0.033\,.
\label{eq:g2NN}
\end{equation}
In this Letter, we will use the numerical framework and the determination of Ref.~\cite{Bacchetta:2025ara} as a baseline to quantify the impact on $g_2$ of lattice data for the CSK in the context of a TMD phenomenological extraction.

\section{Lattice data}
\label{sec:lattice}

Over the last five years, several efforts have been made to extract the CSK from lattice QCD following the original proposals in Refs.~\cite{Ji:2014hxa,Ebert:2018gzl,Ebert:2019okf,Ji:2019sxk,Ji:2019ewn,Ji:2021znw} and follow-up works in Refs.~\cite{Vladimirov:2020ofp,Rodini:2022wic,Deng:2022gzi,Zhao:2023ptv}. Exploratory calculations based on quasi-TMD beam functions have been performed in quenched~\cite{Shanahan:2020zxr} and dynamical~\cite{Shanahan:2021tst,Shu:2023cot} lattice QCD. Analogous calculations based on quasi-TMD wave functions~\cite{Ji:2019sxk,Ji:2021znw} were also carried out~\cite{LatticeParton:2020uhz,LatticePartonLPC:2022eev,Li:2021wvl,Schlemmer:2021aij,Shu:2023cot,Alexandrou:2025xci}.
Recent lattice QCD calculations of the quark CSK using quasi-TMDs in Coulomb gauge~\cite{Bollweg:2024zet,Bollweg:2025iol} have also shown promise for significant reduction in statistical uncertainty. However, at present there is only a single lattice QCD determination of the CSK with full systematic control, including, for the first time, the continuum $a \to 0$ extrapolation~\cite{Avkhadiev:2023poz,Avkhadiev:2024mgd}.

The analysis described in this Letter is thus based on the quark CSK from Refs.~\cite{Avkhadiev:2023poz,Avkhadiev:2024mgd}, which includes three ensembles of gauge-field configurations with lattice spacings $a\in\lbrace a_1=0.15\text{ fm}, a_2=0.12\text{ fm}, a_3=0.09\text{ fm} \rbrace$. This data can be used at two levels of analysis: either extrapolated to the continuum or at finite lattice spacing. The latter level allows us to perform the extrapolation to the physical point {\it simultaneously} with the extraction of $g_2$. This gives us direct control over the modeling of the $b$ dependence of the CSK required to carry out the extrapolation.

The CSK determinations in Ref.~\cite{Avkhadiev:2024mgd} used different parameterizations of $g_K$ than the one employed here. For reference, an analogous fit to the lattice QCD data of Ref.~\cite{Avkhadiev:2024mgd} using the parameterization of Eq.~\eqref{eq:gKparam} gives:
\begin{equation}
    g_2^{\text{Lattice}} = 0.152 \pm 0.027.
\end{equation}

\section{Joint extraction of the CSK}

In this section, we studied the impact on $g_2$ of including the CSK lattice data of Refs.~\cite{Avkhadiev:2024mgd,Avkhadiev:2023poz} into the extraction of Ref.~\cite{Bacchetta:2025ara} following two different strategies: reweighting and simultaneous fit.

\subsection{Reweighting}
\label{s:reweigh}

The impact of incorporating new data into an existing fit can be effectively estimated using the Bayesian reweighting procedure~\cite{Ball:2010gb}. This method consists of assigning a weight to each of the $N=250$ Monte Carlo (MC) replicas of the fit of Ref.~\cite{Bacchetta:2025ara} according to its likelihood with respect to lattice data, which in turn is estimated through $\chi^2$.

We performed the reweighting both on the continuum-extrapolated and on the finite-lattice-spacing data. In the former approach, data for the CSK is provided as a set of $n=21$ points $\{K_j\pm\sigma_j\}$, $j=1,\dots,n$, corresponding to as many values of $b\in\{b_j\}$. Since this data set comes without correlations between points, the $\chi^2$ of the $\alpha$-th replica, with $\alpha=1,\dots,N$ is computed as:
\begin{equation}
\chi^2_\alpha = \sum_{j=1}^{n} \left(\frac{K_j - K^{(\alpha)}(b_j)}{\sigma_j}\right)^2\,,
\end{equation}
where $K^{(\alpha)}(b_j)$ is evaluated using Eq.~(\ref{eq:RGECSKSolNP}) with $g_2=g_2^{(\alpha)}$ extracted from the $\alpha$-th MC replica. The weight to be associated with this replica is~\cite{Ball:2010gb}:
\begin{equation}
w_\alpha = \mathcal{N} \left(\chi^2_\alpha\right)^{n/2-1}e^{-\chi^2_\alpha/2}\,,
\label{eq:weights}
\end{equation}
where $\mathcal{N}$ normalizes the weights so that $\sum_\alpha w_\alpha = N$. Weighted average and variance of $g_2$ over the MC ensemble are then computed respectively as:
\begin{equation}
\langle g_2\rangle=\sum_{\alpha=1}^N w_\alpha g_2^{(\alpha)}\,,\quad\sigma_{g_2}^2=\sum_{\alpha=1}^N w_\alpha(g_2^{(\alpha)}-\langle g_2\rangle)^2\,.
\end{equation}
The result is:
\begin{equation}
g_2^{\rm Rew-cont}\equiv\langle g_2\rangle\pm \sigma_{g_2} = 0.164 \pm 0.020\,.
\label{eq:g2_from_cont}
\end{equation}
The associated number of effective replicas~\cite{Ball:2010gb}, $N_{\rm eff}\simeq150$, indicates that the reweighting was successful. Indeed, $N_{\rm eff}$ is significantly lower than $N$ but still large enough to produce a statistically sound ensemble after reweighting. Comparing the purely phenomenological determination in Eq.~(\ref{eq:g2NN}) with the lattice-reweighted one in Eq.~(\ref{eq:g2_from_cont}), we observe a shift in central value of around 10\% and a reduction in uncertainty of 40\%.

Next, we consider the finite-lattice-spacing
data set. This is subdivided into three subsets corresponding to the lattice spacings discussed above, and counting $n_1=6$, $n_2=7$, and $n_3=8$ points. Each subset $\{K_{j_i}\}$, with $i=1,2,3$ and $j_i=1,\dots,n_i$, comes with its own covariance matrix $\Sigma_i$ estimated in the lattice analysis of Ref.~\cite{Avkhadiev:2024mgd}. Therefore, the $\chi^2$ of the $\alpha$-the replica is evaluated as:
\begin{equation}
\label{eq_chi2_bare_lattice}
\begin{array}{rcl}
\chi^2_\alpha &=&\displaystyle\sum_{i=1}^3 \sum_{j_i, l_i=1}^{n_i} \left(K_{j_i} - \overline{K}^{(\alpha)}(b_{j_i})\right)\\
\\
&\times&\displaystyle(\Sigma^{-1}_i)_{j_i, l_i}\left(K_{l_i} - \overline{K}^{(\alpha)}(b_{l_i})\right)\,,
\end{array}
\end{equation}
and the corresponding weight computed as in Eq.~(\ref{eq:weights}). However, each prediction for the CSK $\overline{K}$ must now correspond to the appropriate lattice spacing $a$. This is done by replacing $g_K$ in Eq.~(\ref{eq:RGECSKSolNP}) with:
\begin{equation}
 g_K(b) = 2g_2^2b^2 - k_{1} \frac{a}{b}- k_{2} \frac{a^2}{b^2}\,.
 \label{eq:gkfinite}
\end{equation}
Since the parameters $k_1$ and $k_2$ were not available from the fit of Ref.~\cite{Bacchetta:2025ara}, we generated a MC ensemble for each of them using as priors the values obtained in Ref.~\cite{Avkhadiev:2024mgd}: $k_1=0.22\pm0.08$ and $k_2=0\pm0.1$.\footnote{Note that $k_2$ was estimated to be negligible in Ref.~\cite{Avkhadiev:2024mgd} based on the Akaike information criterion (AIC)~\cite{AkaikeAIC}. To generate a MC ensemble, we assigned an uncertainty of 0.1 to it.} To consistently include these parameters in the reweighting, we first sampled $10^4$ replicas of $g_2$ uniformly from the original MC ensemble and then generated a pair $(k_1, k_2)$ for each replica according to $k_1 \sim \text{Normal}(0.22, 0.08)$ and $k_2 \sim \text{Normal}(0, 0.1)$.

After reweighting, we find:
\begin{equation}
\label{k1k2g2_from_bare}
\begin{split}
g_2^{{\rm Rew - fin}} &= 0.165 \pm 0.020\,.
\end{split}
\end{equation}
The reweighted value of $g_2$ is in perfect agreement, both in terms of central value and uncertainty, with Eq.~(\ref{eq:g2_from_cont}). We also find that the reweighted values of $k_1$ and $k_2$ agree with their respective priors. This proves the consistency of performing the reweighting using continuum-extrapolated and finite-lattice-spacing lattice data, and provides further evidence of the impact of this data on the CSK. For definiteness, we will take the value in Eq.~(\ref{k1k2g2_from_bare}) as our best estimate of $g_2$ by reweighting.

\subsection{Simultaneous fit}
\label{s:fit}

Following the promising outcome of the reweighting analysis, 
we explore for the first time the impact of CSK lattice data in a TMD fit. Specifically, we performed a fit using the exact same setup of Ref.~\cite{Bacchetta:2025ara}, but including the three sets of finite-lattice-spacing data for the CSK of Refs.~\cite{Avkhadiev:2024mgd,Avkhadiev:2023poz}. Accordingly, predictions are computed using Eq.~(\ref{eq:RGECSKSolNP}) with $g_K$ given in Eq.~(\ref{eq:gkfinite}). We treat these sets by accounting for correlations as encoded in the covariance matrices $\Sigma_i$. In total, we included 503 points in the fit: 482 Drell-Yan (DY) experimental points and 21 lattice points. In the fit we used a training-validation split of 50\%-50\% for experimental data. However, due to its limited amount, lattice data is fully included in the training set.

In Tab.~\ref{t:chi2}, we compare the
quality of the baseline fit of Ref.~\cite{Bacchetta:2025ara} with the reweighted results  and the combined fit of experimental and lattice data. For each data subset (fixed-target, RHIC, Tevatron, LHCb, CMS, ATLAS, and lattice) we list the number of points included in the fit ($N_{\rm dat}$), and  the $\chi^2$ averaged over the MC ensemble, $\langle\chi^2\rangle$.\footnote{Differently from Ref.~\cite{Bacchetta:2025ara} where we quoted the reduced $\chi^2$'s of the central replica, here we report $\langle\chi^2\rangle$ because no central replica can be defined in the case of reweighting. The $\langle\chi^2\rangle$'s in the reweighting case are to be understood as weighted averages.}

\begin{table}[h]
    \centering
	\begin{tabular}{lcccc}
		\hline
	   \vphantom{\Big|}Experiment\quad & $N_{\text{dat}}$ & \multicolumn{3}{c}{$\langle\chi^2\rangle$}\\
		\vphantom{\Big|}                &                  & Baseline  & 
        Reweighting &  Fit\\ 
        \hline
		\vphantom{\Big|} Fixed-target   & 233              & $239.67$  & $234.09$ & $242.29$ \\
		\vphantom{\Big|} RHIC           & 7                & $7.06$  & $7.11$ & $7.62$ \\
		\vphantom{\Big|} Tevatron       & 71               & $61.33$  & $59.91$ & $65.31$ \\
		\vphantom{\Big|} LHCb           & 21               & $22.94$  & $21.88$ & $22.77$ \\
		\vphantom{\Big|} CMS            & 78               & $30.74$  & $29.85$ & $29.48$ \\
		\vphantom{\Big|} ATLAS          & 72               & $95.35$  & $95.60$ & $95.71$ \\
        \hline
        \vphantom{\Big|} Lattice $a_1$      & 6           & \textit{31.13}  & $11.5$ & $12.56$ \\
        \vphantom{\Big|} Lattice $a_2$      & 7           & \textit{6.30}  & $4.67$ & $1.77$ \\
        \vphantom{\Big|} Lattice $a_3$      & 8           & \textit{8.50}  & $3.88$ & $6.01$ \\ 
		\end{tabular}
	\caption{Values of the $\langle\chi^2\rangle$'s for
	each data subset, along with the respective numbers of points $N_{\rm dat}$, obtained with
	baseline fit in Ref.~\cite{Bacchetta:2025ara}, reweighting, and simultaneous fit. The $\langle\chi^2\rangle$'s of lattice data obtained using the baseline fit, which does not include this data, are shown in italic.}
	\label{t:chi2}
\end{table}

We observe that the average $\langle\chi^2\rangle$ values for the DY data are stable across the three configurations. This indicates that there is no tension between experimental and lattice data in that the inclusion of the latter does not cause any deterioration in the description of the former.

It may be argued that the reduced number of lattice points can hardly affect the description of the much more abundant experimental data set. Although this is true, the marked improvement in the description of lattice data upon inclusion in the fit (or by reweighting) demonstrates their constraining power.

The value of $g_2$ obtained from the simultaneous fit is:
\begin{equation}
    g_2^{\rm Fit} = 0.167 \pm 0.015\,.
    \label{e:g2_fit}
\end{equation}
This result is in very good agreement with the value obtained through reweighting in Eq.~(\ref{k1k2g2_from_bare}), both in terms of central value and uncertainty. 
\begin{figure}
    \centering
    \includegraphics[width=1.0\linewidth]{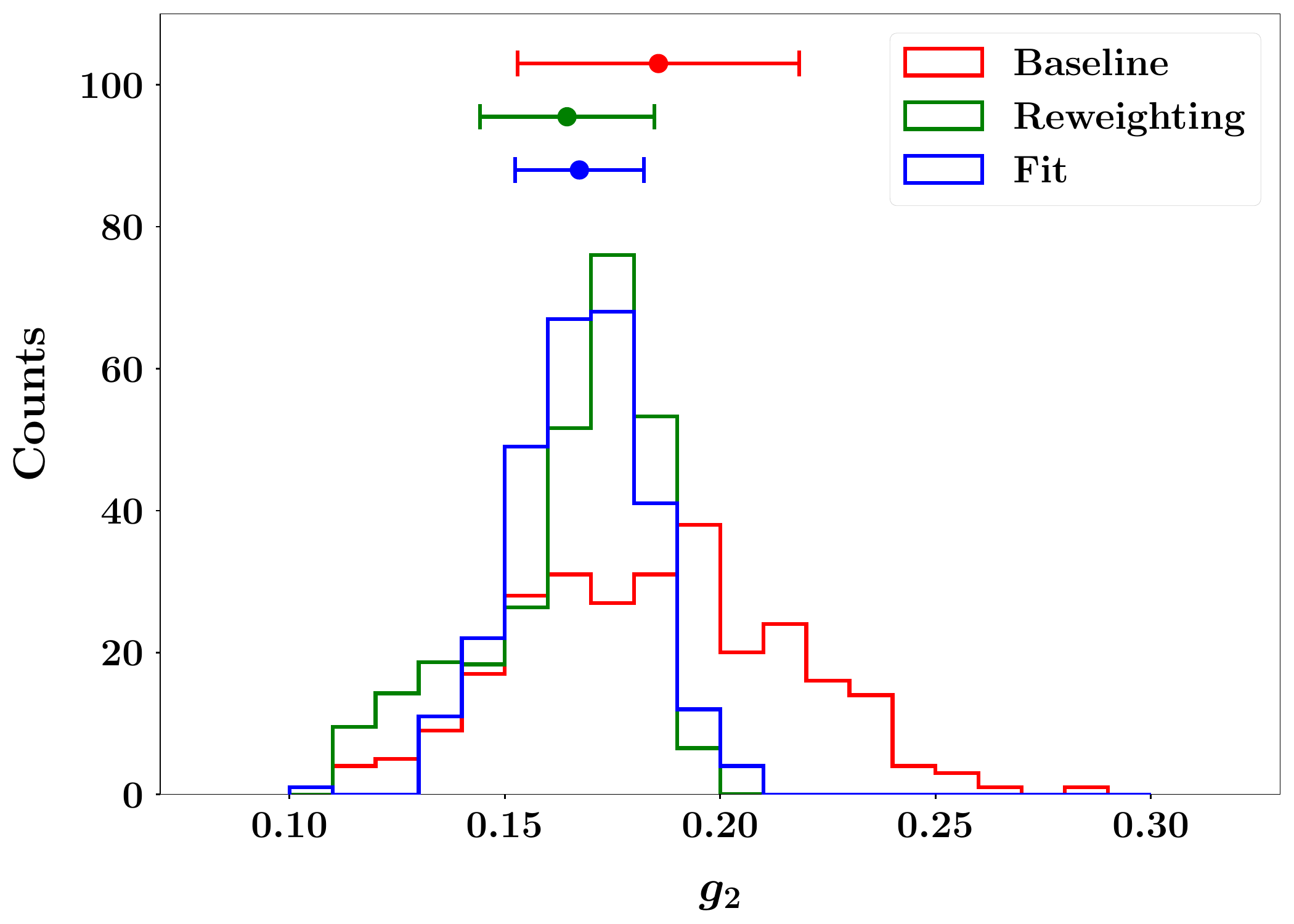}
    \caption{MC replica distribution of $g_2$ from the baseline fit (red), after reweighting (green), and from the simultaneous fit (blue).
    The points above the distributions display the respective central values and one-$\sigma$ uncertainties.}
    \label{fig_g2_fit_histogram}
\end{figure}
A quantitative comparison is given in Fig.~\ref{fig_g2_fit_histogram}, where the distributions of $g_2$ over the respective MC ensembles for baseline fit, reweighting, and simultaneous fit are compared. This plot shows the consistency of reweighting and simultaneous-fit determinations, with the latter having a slightly smaller uncertainty. Moreover, shift and uncertainty reduction on $g_2$ caused by the inclusion of lattice data are evident when comparing reweighting and simultaneous-fit determinations to the baseline one.

\begin{figure}
    \centering
    \includegraphics[width=1.0\linewidth]{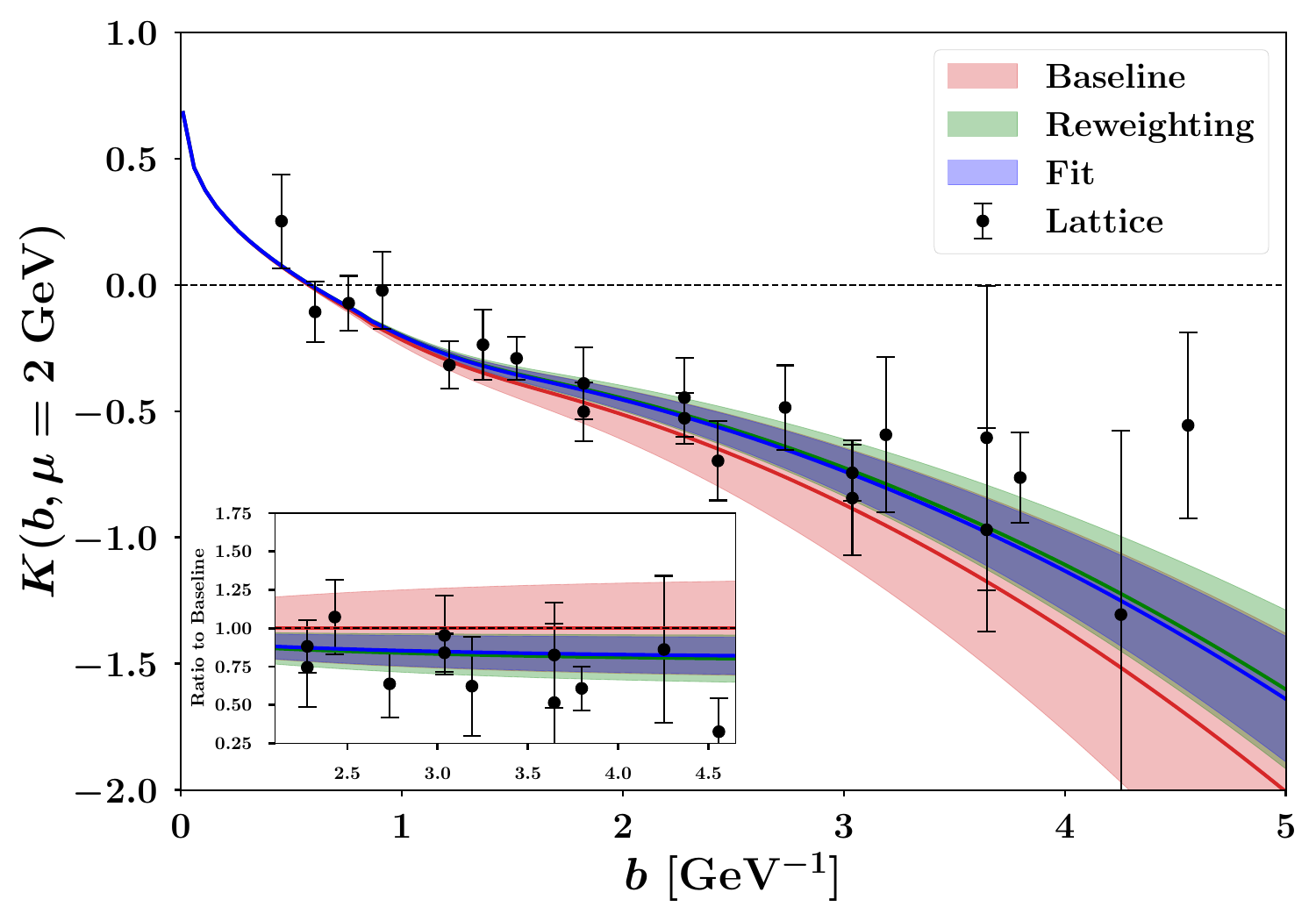}
    \caption{The CSK as a function of $b$ at $\mu= 2$~GeV as obtained from the baseline of fit of Ref.~\cite{Bacchetta:2025ara} (red band), from  reweighting (green band), and from the simultaneous fit of experimental and lattice data (blue band). Bands correspond to one-$\sigma$ uncertainties. Data points correspond to the extraction of Ref.~\cite{Avkhadiev:2024mgd}. The bottom inset shows the ratio to the baseline predictions at large values of $b$.}
    \label{fig:CS_comparison}
\end{figure}
Fig.~\ref{fig:CS_comparison} displays the CSK computed as in Eq.~(\ref{eq:RGECSKSolNP}), with $g_K$ given in Eq.~(\ref{eq:gKparam}), as a function of $b$ at the scale $\mu= 2$~GeV. Curves corresponding to baseline fit, reweighting, and simultaneous fit are shown, along with the continuum-extrapolated lattice data shown as black points. This plot confirms the consistency of reweighting and simultaneous fit, as well as the significant impact of lattice data on the CSK. 
This is made particularly clear by the bottom inset of Fig.~\ref{fig:CS_comparison}, where the ratio to the baseline predictions at large values of $b$ is shown. Indeed, the CSK in this region is maximally sensitive to the nonperturbative parameter $g_2$, while the small-$b$ region is mostly determined by the perturbative components.

We conclude this section by noting that, although lattice data for the CSK has a strong impact on $g_2$ when included in a TMD determination, the resulting TMD PDFs remain very stable.

\section{Conclusion}

We presented the first successful joint extraction of the nonperturbative Collins-Soper kernel (CSK) by combining experimental Drell-Yan data with lattice QCD calculations in the context of a TMD analysis based on a neural-network paratrization. The inclusion of lattice data consistently shifts the central value of the parameter $g_2$, which governs the behavior of the CSK in the nonperturbative region, by approximately 10\% and reduces its uncertainty by 50\%, leading to a significantly more precise determination. Two independent methodologies\thinspace---\thinspace Bayesian reweighting of an existing fit and a simultaneous fit\thinspace---\thinspace yield remarkably consistent results, demonstrating the robustness of the approach. The analysis finds no tension between experimental and lattice data, and the description of the Drell-Yan data remains stable. This study highlights the significant potential of lattice QCD inputs to substantially improve precision in the extraction of transverse-momentum-dependent distributions. 

\section{Acknowledgments}
The work of S.R. is supported by the German Science Foundation (DFG), grant
number 409651613 (Research Unit FOR 2926), subproject 430915355. The work of V.B. has been supported by l’Agence Nationale de la Recherche
(ANR), project ANR-24-CE31-7061-01. This material is based upon work supported by the U.S. Department of Energy (DOE), Office of Science, Office of Nuclear Physics through Contract No.~DE-AC02-06CH11357.
Argonne National Laboratory’s contribution is based upon work supported by Laboratory Directed Research and Development (LDRD) funding from Argonne National Laboratory, provided by the Director, Office of Science, of the U.S. DOE under Contract No. DE-AC02-06CH11357.
This manuscript has been authored by FermiForward Discovery Group, LLC under Contract No. 89243024CSC000002 with the U.S. Department of Energy, Office of Science, Office of High Energy Physics.
A.A., Y.F., and P.E.S. were supported in part by the U.S.~Department of Energy, Office of Science, Office of Nuclear Physics, under grant Contract Number DE-SC0011090, by Early Career Award DE-SC0021006, by Simons Foundation grant 994314 (Simons Collaboration on Confinement and QCD Strings), by the U.S. Department of Energy SciDAC5 award DE-SC0023116, and have benefited from the QGT Topical Collaboration DE-SC0023646.
V.B., A.A., and P.E.S thank the Department of U.S. DOE Institute for Nuclear Theory (INT) at the University of Washington for its hospitality and the Department of Energy for partial support during the completion of this work; this research was supported in part by the INT's U.S. DOE grant No. DE-FG02- 00ER41132.

\bibliographystyle{myrevtex}
\bibliography{biblio}
\end{document}